\documentclass{aa}
\usepackage{graphicx}
\usepackage{txfonts}
\begin{document}
\title{The hard to soft spectral transition in LMXBs - affected by
  recondensation of gas into an inner disk}


\author{E. Meyer-Hofmeister \inst{1}, B.F. Liu\inst{2} and F. Meyer\inst{1}}
\offprints{Emmi Meyer-Hofmeister; emm@mpa-garching.mpg.de}
\institute
    {Max-Planck-Institut f\"ur Astrophysik, Karl-
     Schwarzschildstr.~1, D-85740 Garching, Germany\\
     \email{emm@mpa-garching.mpg.de}
\and
    National Astronomical Observatories/Yunnan Observatory, Chinese
    Academy of Sciences, P.O. Box 110, Kunming 650011, China\\
     \email{bfliu@ynao.ac.cn}}

\date{Received: / Accepted:}
\abstract
   {Soft and hard spectral states of X-ray transient sources reflect
    two modes of accretion, accretion via a geometrically thin,
    optically thick disk or an advection-dominated accretion flow (ADAF).}
  {The luminosity at transition between these two states seems to vary 
    from source to source, or
    even for the same source during different outbursts, as observed
    for GX 339-4. We investigate how the existence of an inner weak
    disk in the hard state affects the transition luminosity.}
    {We evaluate the structure of the corona above an outer truncated
    disk and the resulting disk evaporation rate for different irradiation.}
  {In some cases, recent observations of X-ray transients indicate the
    presence of an inner cool disk during the hard state. Such a disk
    can remain during
    quiescence after the last outburst as long as the luminosity does
    not drop to very low values ($10^{-4}$-$10^{-3}$ of the Eddington
    luminosity). Consequently, as part of the matter accretes via the
    inner disk, the hard irradiation is reduced. The hard irradiation
    is further reduced, occulted and partly reflected by the inner disk. This
    leads to a hard-soft transition at a lower luminosity.}
  {The spectral state transition is expected at lower luminosity
  if an inner disk exists below the ADAF. This seems to be supported
  by observations for GX 339-4.}



\titlerunning {
The spectral transition - affected by an inner cool disk
}

\maketitle
%

\section{Introduction}

Accretion onto black holes is an interesting phenomenon since the
main features are similar for stellar mass and supermassive black
holes. Concerning both groups of sources, different
solutions of the hydrodynamic equations of viscous
differentially rotating flows are known (for a discussion see Narayan
et al. 1998). The two common modes of accretion are either
via a geometrically thin, optically thick disk reaching
inward to the last stable orbit (for accretion rates between a few
percent up to almost the Eddington rate), or, via an outer disk
truncated some distance from the center, and a hot optically
thin, geometrically extended advection-dominated flow (ADAF) in the
inner region. This latter type of accretion flow, present for
 accretion rates below a few percent of the Eddington rate, was mainly
investigated by Narayan and
collaborators (Narayan \& Yi 1994, 1995a, 1995b and Abramowicz et
al. 1995). If the mass accretion rates  and the radii are scaled to
the Eddington
accretion rate ($\dot M_{\rm{Edd}}=L_{\rm{Edd}}/\eta c^2, L_{\rm{Edd}}=4\pi
GMc/ \kappa, \kappa$ electron scattering opacity, $\eta$ radiation efficiency) and Schwarzschild radius
($R_S=2GM/c^2$) the
basic equations are the same for small and large
black holes and for neutron stars. These two  main accretion regimes
are present in low-mass X-ray binaries (LMXBs)
and active galactic nuclei (AGN).

New data of X-ray properties of black holes in AGN from the
{\it{Chandra}} X-ray Observatory allow us to study the accretion
process also for  AGN of low luminosity in more detail, especially
galactic nuclei with an ADAF near the center (for a review see Ho
2008). The galaxies can have different nuclear activity ranging from
luminous AGN to the least active galaxies such as our Galaxy (for a
review see Yuan 2007). To understand the
physical processes in supermassive black hole accretion, one can make
use of the knowledge gained from studying low-mass X-ray binaries
(LMXBs). For these systems, spectral state transitions and other detailed
features can be observed and studied. The innermost 
regime around the black hole also is of interest as relativistic
features can allow the determination of black hole spin
(Miller 2007, Miller et al. 2009).

Our investigation concerns the transition from a hard spectral state
to a soft spectral state due to an increase of the mass flow rate in the
accretion disk as happens during the rise to outburst. The now commonly
accepted picture for the state transition is related to
the change of the accretion mode in the inner region from an ADAF (with a hard
spectrum originating from the very hot ADAF) to a thin disk
extending inward to the last stable orbit (with a soft black body spectrum).
For an updated account of the ADAF model see Narayan \&
McClintock (2008).

Fits confirming truncated disks with inner ADAF-filled regions were
worked out for LMXBs, first for A0620-00 and V404 Cyg (Narayan et
al. 1996, 1997).  Esin (1997) applied the model to
state transitions observed for SXT Nova Muscae and Cyg X-1, leading to
a description of the configuration of the accretion geometry for
different spectral states that varies with the mass flow rate (Esin et
al. 1997), adequate for black holes of
different mass. Applications to Galactic nuclei were equally
successful; the first application concerned Sgr A* and successfully
explained its low radiative efficiency (Narayan et al. 1995c). For a
review of early investigations of AGN see Narayan et al. (1998).

The investigation of the interaction between a coronal flow and
an outer accretion disk led to the result that heat
conduction causes evaporation of matter from the disk to the hot flow
(Meyer et al. 2000a, Liu et al. 2002). An equilibrium establishes
between the cool disk flow and the hot flow. The balance between
disk mass accretion rate and evaporation rate determines where
the disk flow changes to an ADAF. This process explains the location
of the disk truncation as well as the dependence of the spectral state
transition on the accretion rate (Meyer et al. 2000b).

Over the past several years, a large number of observations has become
available. Current X-ray missions (in
particular {\it{Chandra}} and {\it{XMM-Newton}}) have a sensitivity
that allows us to study the accretion flow, ADAF or
disk flow, as well as the transitions between them, in detail. Most 
sources were detected in the high luminosity, soft spectral state. If
in outburst the rise and the decline phase
could be observed, and possibly both spectral state transitions, it was
found that the luminosity at hard/soft transition during the rise to
outburst usually was higher  than the one at the soft/hard transition during
outburst decline, which means that a source can be in either hard or soft state
at a certain luminosity. This so-called hysteresis was
first found by Miyamoto (1995) in observations of GX 339-4. The
luminosity difference between the two transitions is about a factor of
3-5. Theoretically this hysteresis
could be understood as caused by the different type of irradiation of the corona
during hard and soft spectral states (Meyer-Hofmeister et al. 2005).
Hardness-intensity diagrams (HID), now deduced for a number of
sources, show this hysteresis clearly, together with other features
such as the times at which quasi-periodic oscillations (QPO) and a
jet and radio emission appeared.

Observations seem to indicate that the luminosity at which the hard/soft
spectral transition occurs can differ, either from source to source
(a comparison is difficult unless the luminosities are scaled to the Eddington
luminosity) or even for the same source from outburst to outburst, as in
the black hole transient GX 339-4. Observations are discussed by Dunn
et al. (2008); hardness-intensity
diagrams of four outbursts clearly show the transition at varying
luminosity. Hard/soft state transitions of GX 339-4 were
also investigated by Belloni et al. (2006), Nowak (2006),
Caballero-Garc\`ia et al. (2009) and Del Santo et
al. (2009). Why the hard/soft transitions for black hole sources and possibly
also for neutron star transients can occur at different
luminosity is still an open question.

Already from RXTE observations some evidence was found for soft
components in the low/hard state for a number of sources. Recently 
{\it{XMM-Newton}} observations clearly revealed the presence of disk
material very close to the innermost circular orbit in
GX 339-4 and SWIFT J1753.5-0127 (Miller et al 2006a,b).
It is the aim of our paper to investigate
whether an inner cool disk underneath the ADAF existing during the
hard spectral state can affect the state transition. In Sect. 2 and 3
we briefly describe the phenomena of disk truncation, formation of a
gap and a leftover weak inner disk. We summarize the observational
evidence for an inner disk in Table 1. In Sect. 4 we discuss
hard and soft irradiation of the corona by the ADAF and the inner
disk. The hard irradiation of the corona is reduced if part of the
matter accretes via the inner disk instead of via the ADAF. The inner disk
causes further changes, e.g. less hard irradiation due to occultation of
the ADAF and reflection. This results is a spectral state transition
at lower luminosity than in the case of no inner disk (Sect. 5). In
Sect. 6 we compare with observational evidence in different sources. In Sect. 7 we consider the hardness-intensity
diagram (HID) of GX 339-4. A discussion and conclusions follow.

\section{Disk truncation}
\subsection{Disk truncation during the low luminosity state}
For our investigation of the influence of an inner disk on the
spectral state transition, the accretion history of a transient source,
especially truncation of the disk and the formation of
a gap between outer and inner disk, is essential. 

ASM aboard {\it{RXTE}} has provided a continuous,
seven-year record of the activity of many galactic sources. In the
review of McClintock \& Remillard (2006), light curves of black hole
binaries, or black hole candidate binaries, are shown together with the
changes of the hardness ratio defined as the ratio HR2 of counts in
the 5-12 keV and the 3-5 keV range. The outbursts typically begin with
and end in
the low/hard state  when the disk is truncated (for GS/GRS 1124-68 and
XTE J1550-564 changes of the inner edge of the disk are shown). Remillard \&
McClintock (2006) review the X-ray states of 20 X-ray binaries.
Yuan \& Narayan (2004, Fig.3) present the accretion luminosity and
disk truncation radius for AGN and LMXRBs. For a low mass accretion
rate, i.e. a low luminosity below $10^{-6}L_{\rm{Edd}}$, the disk is truncated far
outside, at distances greater than $10^4$ Schwarzschild radii.

Recent {\it{Chandra}} observations detected two black
hole sources, XTE J1550-564 and H1743-322, at their faintest level of
X-ray emission (Corbel et al. 2006, Bradley et al. 2007). As a new
feature of the accretion geometry of truncated disks, for some sources
the {\it  XMM-Newton } spectra reveal the presence
of a cool accretion disk component and a relativistic Fe K emission
line (Miller et al. 2006b), see Sect. 3.1. One might ask
whether these results, which
show a thermal component and an ADAF coexisting in
the innermost region, cast doubts on the commonly accepted truncated
disk picture. However, recent theoretical work (Liu et
al. 2006, Meyer et al. 2007) showed that
a weak inner disk can be sustained as a consequence of gas
condensation from the ADAF downward into a cool disk.

\subsection{The disk truncation
  during the rise to an outburst and the decline}
In their recent review Remillard \& McClintock (2006) gave a
quantitative three-state description of active accretion. This
description mainly uses the ratio of the disk flux to total flux, the
power-law photon index $\Gamma$, the continuum power in the power
density spectrum and the occurrence of quasi-periodic oscillations
. This new nomenclature and the new definitions of
spectral states were intended to provide a better ordering of spectral
states.

The sharpest point of departure from the previously used definition of
states, as the authors call it (McClintock \& Remillard 2006), is
that luminosity is abandoned as a criterion for defining the state of
the source. The hysteresis in the luminosity of state transition seems to
suggest that we not use the luminosity as an important parameter. But this is
only a limited luminosity interval where the bi-modality due to
different irradiation from the differently structured innermost region exists
(as described later). Narayan \& McClintock (2008) take
the mass accretion rate as the key parameter that mainly determines
the spectral state of an accreting black hole.
 
The truncation of the geometrically
thin, optically thick Shakura-Sunyaev disk at a certain distance is
caused by the interaction of the corona and disk. In the model
proposed (Meyer et al. 2000a) the corona is fed by matter
which evaporates from the cool layers of the disk underneath. Since in
the closer regions evaporation is very efficient, for low accretion rates all
matter is transferred to the corona (at the distance where accretion
rate equals the evaporation rate). The gas proceeds towards the black hole
as a purely coronal vertically extended flow. While in quiescence in
LMXBs the disk is truncated, during the rise to an outburst the
increasing mass flow rate in the disk, in most cases, becomes larger
than the maximal evaporation rate, so that the disk is not truncated
anymore. Then, within the diffusion time, which is short, of the order of
hours to days for different distances, the inner disk region is filled
with an accretion disk and the spectrum becomes soft.

\begin{figure}
\includegraphics[width=8.0cm]{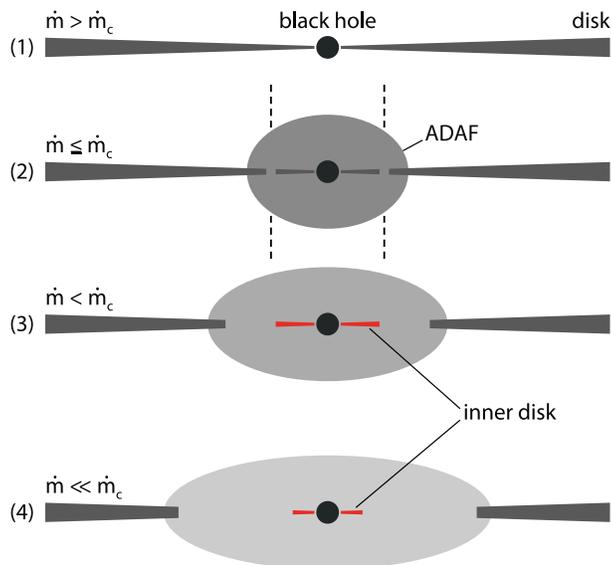}
\caption{\label{schematic}Geometry of the accretion flow as a function
  of the mass accretion rate scaled to the Eddington rate $\dot m$,
  ($\dot m_c$ critical rate for which the state transition happens. (1) to
  (4) describes the change from a soft state with high mass
  flow rate, to the formation of a gap separating the outer from the
  inner disk, an inner disk, beginning of
  the hard state. The leftover inner disk becomes weak, but can
  remain as long as the mass accretion rate does not drop below about
  $10^{-3}-10^{-4} \dot M_{\rm{Edd}}$. The intensity of
  gray color indicates the mass flow rate in the ADAF (compare Narayan
  \& McClintock 2008).}
\end{figure}

The reverse process happens when during the decline from an
outburst the accretion rate decreases (Meyer et al. 2007).
Fig. \ref{schematic} describes this process: in state (1) the disk
reaches inward for high accretion rates; state (2) is reached if the accretion
rate becomes lower than
the maximal evaporation rate $\dot m_c$, which is a critical value for the
accretion geometry; a gap in the disk appears at that
distance that will be filled with an ADAF; state (3): with further
decrease of the accretion rate the gap widens and
the spectrum becomes hard; n state (4) the inner disk remains as long as
re-condensation of matter from the
ADAF allows (see Sect.3.2). We investigate
whether this leftover inner disk affects the hard/soft spectral state
transition.

\begin{table*}[t]
\begin{center}
\caption{\label{t:fits} 
\large{\bf{Observations: Inner disks in the
  low/hard state }}          }
\begin{tabular*}{155mm}[t]{@{\extracolsep{\fill}}llllcl}
\hline
&source & outburst phase & thermal component & inner disk ?
& references\\
&  & &  $L_{\rm{disk}}/L_{\rm{Edd}}$ & & \\
\hline
& &&&& \\
& GX 339-4 & outburst 2004, rise & $5\,10^{-2}$ &  present& 1,2\\
&GX 339-4 & outburst 2007, decline & $2.3\,10^{-2} $ &  present &3\\
&GX 339-4 & outburst 2007, decline & $8\,10^{-3} $
  & present (weaker evidence) & 3\\
&J1753.5-0127 & outburst 2006, decline & $3\,10^{-3} $ &
present &4\\
&XTE J1817-330 & outburst 2006, decline & $1.2\,10^{-3} $
& present &5\\
&XTE J1118+480 &  hard state outburst 2000 &$10^{-3} $   & present & 6,7\\
&GRO J1655-40 & outburst 2005, rise  &  $10^{-3}$  & small
significant evidence &5\\
&V4641 SGR & outburst 1999, rise & $1.5\,10^{-3}$ ?&present &8\\
&XTE J1650-500 &outburst 2002, rise ? & $10^{-3}$ &present & 9,10\\
& & outburst 2002, late decline &$ \le 10^{-4}$ & no evidence &11\\
&4U 1543-47 & outburst 2002, late decline  & 
  $10^{-5}-10^{-6}$  &  some evidence &12,13\\
&Cygnus X-1  & hard intermediate state 2001&$7\,10^{-2}$ & present&
10,14\\
&SAX J1711.6-3808 &outburst 2001, decline & .\,\,.\,\,. &present  ? & 15\\
&XTE J1908+094 &outburst 2002 & .\,\,.\,\,.&present ? & 16\\
&V404 Cyg   & quiescence & $\le 10^{-6}$ &no evidence &17\\
&XTE 1550 &&&&\\
&and H1743-322 & quiescence & $\le 10^{-6}$ & no evidence& 18\\
\hline

\end{tabular*}
\end{center}

References: 1) Miller et al. 2006b, 2) Reis et al. 2008, 3) Tomsick et
al. 2008, 4) Miller et al. 2006a, 5) Rykoff et al. 2007, 6)
McClintock et al. 2001, 7)  Reis et al. 2009, 8) Miller et al. 2002c,
9) Miller et al. 2002a, 10) Miller et al. 2009, 11) Tomsick et
al. 2004,  12) Kalemci et al. 2005, 13) La Palombara \& Mereghetti
2005, 14) Miller et al. 2002b, 15) in't Zand et al. 2002a, 16) in't
Zand et al. 2002b, 17)  Bradley et al. 2007, 18) Corbel et al. 2006. 

For a discussion of details and sources with peculiar dim soft
  states, GRS 1758-258 and 1E 1740.7-2942, see text.
\end{table*}

\section{An inner cool disk during the hard state ?}
\subsection {Observations}

For several sources, the presence of an inner disk during the low/hard
state, mostly in its brightest phases (Tomsick et
al. 2008), seems indicated. Evidence was already found
from {\it{RXTE}} observations and {\it{XMM-Newton}} spectra now allow
us to
search for a soft component at very low luminosity. However there are
still few observations where both the presence of an inner disk is
indicated and also the 
luminosity at a subsequent hard/soft transition (the specific topic of
our analysis) is known. But the presence of an inner disk in a
number of sources shows that the spectral transition might often
be affected.
In Table 1 we summarize observations that show a soft
component during the low/hard spectral state, indicating the presence
of a weak accretion disk close to the last stable orbit, 
surrounded by a more spherically extended ADAF. In the following we
give more detailed information on the listed sources. The estimates
for the luminosity of the soft
component lie in a wide range from a few percent of the Eddington
luminosity down to almost a factor of 1000 lower.

Interesting results were found especially for GRS 339-4 by Miller et
al. (2006b), Reis et al. (2008) and Tomsick et al. (2008). Soft
components were  also clearly revealed for SWIFT J1753.5-012 by Miller et
al. (2006a) and for XTE J1817-330 by Rykoff et al. (2007). Some
evidence for a soft component was also found for XTE J1118+480 in
April 2000, shortly after the discovery of the source when its
luminosity was near 0.001 $L_{\rm{Edd}}$ (McClintock et al. 2001), 
confirmed in recent work (Reis et al. 2009). For GRO J1655-40 the
evidence is small, but significant. V4641 SGR appears to have an
inner disk during the rise to the otherwise untypical outburst
1999 (Miller et al. 2002c). For 4U 1543-475 indications of an inner
disk were found in the decline from the 2002 outburst during the hard
intermediate state (Park et al. 2004)) and during the hard/low
state (Kalemci 2005, La Palombara \& Mereghetti 2005). For XTE J1650-500 an 
{\it{XMM-Newton}} observation early in the 2001 outburst 
showed indications of an inner
disk; the spectral state was first classified as `very high'' (Miller et
al. 2002a), later as low/hard (Rossi et al. 2005, Miller et
al. 2009). During  the late outburst decline of the source at
luminosity levels of $10^{-4} L_{\rm{Edd}}$, Tomsick et al. (2004) did
not find indications for an inner disk.

Such thermal components in the spectra were found for both outburst
rise and/or decline. For our investigation, it is of special interest
whether during the rise to an outburst an inner disk is indicated. This
would mean that such a disk had not disappeared
completely since the last outburst. 

Cygnus X-1 is a special case as the system is persistently active with
only very moderate luminosity changes and it might have an inner disk
close to the innermost stable orbit during bright phases of the
low-hard state (Miller et al. 2002b). Two other 
systems, GRS 1758-258 and 1E 1740.7-2942, also have only small
luminosity variations in a hard spectral state, but show
abrupt transitions to a lower luminosity, a dim soft
state (Pottschmidt et al. 2006, del Santo et al. 2005). The cool disk 
emission in these low flux phases seems to be of a different nature. 

In addition, inner disks might be present in SAX J1711.6-3808 and XTE
J1908+094 (in't Zand et al. 2002a, 2002b), where iron lines indicate an
inner disk, but a disk continuum is not visible, possibly 
due to high line of sight absorption which could easily hide a cool
disk (Miller et al. 2006b). These sources
are strong black hole candidates.

For several of the sources listed in Table 1, further information can 
be found in the work of Miller et al. (2009) on black hole spin parameters.

Exploring the presence of a soft component during the hard spectral
state at very low accretion rates requires long observations. 
For some systems in
quiescence at very low luminosity, $L/L_{\rm{Edd}} \le 10^{-6}$, high
quality spectra did not show evidence of a thermal component or iron
line features, e.g. Bradley et al. (2007) for V404 Cyg,  Corbel et
al. (2006) for XTE 1550 and H1743-32. 

\subsection{Theoretical interpretation}

So far, detections and non-detections point to the existence of an
inner disk, in some sources, and during limited time intervals of
the outburst cycles. What determines whether an inner disk can be
present? As recently found, matter can recondense from the ADAF to an
underlying inner disk (Liu et al. 2006, Meyer et al. 2007) and
therefore sustain an inner disk in the equatorial plane,
left over from the last soft state (without re-condensation, the disk
would disappear within the diffusion time of a few days). The
re-condensation model was applied to GX 339-4 and SWIFT J1753.5-0127
(Liu et al. 2007).

A recent detailed analysis of the observations of
GX 339-4 (Taam et al. 2008) took into account conductive cooling of the
coronal gas  and Compton cooling to evaluate
the temperature of the thermal component at the observed
luminosities. The result of these investigations
is that generally an inner disk can exist for
luminosities in the range of about 0.001 to 0.02 $L_{\rm{Edd}}$.
Such inner disks are weak and usually have a small extension,
from almost the last stable orbit outward to a distance depending on
the mass flow rate in the ADAF, e.g. to around 100 $R_S$ as found in the
analysis of the observations for GX 339-4 (Taam et al. 2008). Between
the inner disk and the outer truncated disk is a gap, an ADAF fills
the region between the truncation of the outer disk and the center, and
surrounds the inner disk (see Fig.\ref {schematic}).
Since the inner disk can exist for a long time during the low
hard state, as long as the mass accretion rate does not become very
low, an inner disk is expected whenever a source does not become very
faint in quiescence (e.g. indicated for GX 339-4 (2004) and GRO
J1655). This means, that not only the mass flow rate at a
certain time during the outburst cycle determines the accretion flow
geometry, it also depends on the mass flow history as to whether an inner
disk can be left over. Usually there are not enough observations
during quiescence to know how low the accretion rate became.

\section{An inner disk below the ADAF}
\subsection{The vertical structure of the corona}
At low luminosities the process of accretion in LMXRBs includes an
inner ADAF region and an outer, geometrically thin optically thick
disk. At the innermost region, within a certain distance
the cool disk and the corona/ADAF exist
together. The interaction can go in two ways, depending on the distance
from the center (this is related to whether radiative loss or advective
cooling dominates, Meyer et al 2000a). Outside a critical distance
the interaction causes evaporation of gas from the disk into the hot
flow, and interior gas from the ADAF condenses into a disk.
Evaporation and condensation are affected by irradiation from the central region.

The vertical structure of the irradiated corona has to be evaluated to find the
maximal evaporation rate. In the inner region, the flow accretes
 partially via an inner ADAF and partially via an inner disk.
In this analysis, the irradiation from the central ADAF and the additional irradiation/Compton cooling from the inner disk are taken into account.

\subsection{Compton heating/cooling due to an inner disk}
During the hard spectral state, irradiation of the
corona originates from different sources: (1) hard radiation from the hot ADAF
filling the inner region, (2) soft radiation from the
truncated disk underneath, and (3) if an inner cool disk exists, additional soft
radiation from this (usually weak) inner disk. The Compton heating/cooling rate per unit volume caused by the hard
radiation from the inner region filled with hot gas of the ADAF is
\begin{equation}\label{compt1}
 {q_{\rm Comp}(z)}^{\rm{ADAF}}=\frac{4k T_e(z)-h\bar{\nu}}{m_e c^2}n_e(z)\sigma_T F_{c}
\end{equation}
with $k$ the Boltzmann constant, $T_e(z)$ electron temperature, $h\bar{\nu}$
mean photon energy, $m_e$ electron mass, $c$ velocity of light,
$n_e(z)$ electron number density, $\sigma_T$ Thomson cross
section and $ F_{c}$  the irradiating flux density. $ F_{c}$ is determined by the luminosity $L_c$ and hence by the central accretion rate via ADAF, $\dot M_c$,
\begin{equation}\label{compt1a}
 F_c={L_c\over 4\pi R^2}, \hspace{0.3cm}
 L_c=\eta_c \dot M_c c^2
\end{equation}
with $R$ the distance from the center and $\eta_c$ the energy conversion efficiency of the corona, which could be less than 0.1.
The mean photon energy $h\bar{\nu}$ essentially determines
the contribution. The effect of hardness of the irradiation from the hot ADAF
gas was investigated in earlier work  in connection with
hysteresis in LMXBs (Liu et al. 2005).

If a (weak) inner disk is present, due to re-condensation of gas from
the corona to the disk, a further contribution to the Compton
cooling/heating process comes from the interaction of photons of this
inner disk with the electrons in the corona.
\begin{equation}\label{compt2}
 q_{\rm {Comp}}(z)^{\rm{inner disk}}=\frac{4k T_e(z)}{m_ec^2} n_e(z)\sigma_T F_{\rm{disk}}
\end{equation}
with $ F_{\rm{disk}}$ the disk flux.
The inner disk appears under an inclination dependent on the
height $z$ so that
\begin{equation}\label{compt2a}
 F_{\rm{disk}}=\frac{L_d}{4\pi R^2} \frac{2z}{(R^2+z^2)^{1/2}},
   \hspace{0.3cm} L_d=0.1\dot M_dc^2
\end{equation}
 with $L_d$ the accretion rate via the disk.

In Eq.\ref{compt2} the term $h\bar{\nu}$ can be neglected
since there $h\bar{\nu} \ll 4k  T_e(z)$. Note that the presence of the inner
disk also means that simultaneously the flux through the corona is
reduced since part of the coronal flow condenses and accretes through the disk.

\begin{figure}[h]
\includegraphics[width=8.cm]{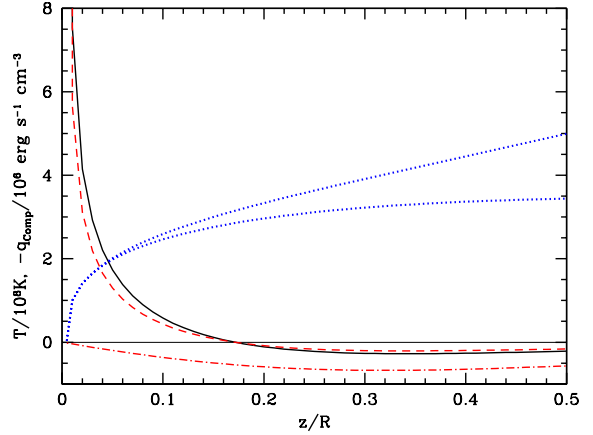}
\caption{\label{Compton} Distribution of Compton heating ($q_{\rm Comp}(z)<0$)
  and cooling ($q_{\rm Comp}(z)>0$) in the corona, illustrated for a model of
  maximal evaporation rate $\dot m=0.029 \dot M_{\rm{Edd}}$
  ($R=1100R_S$, $h\bar{\nu}$= 100 keV, $\eta_c$=0.05).
  Solid black line: Compton heating/cooling caused by
  irradiation from the hot inner ADAF, standard case; dashed red line: Compton
  heating/cooling for reduced hard irradiation (75\% of the total
  accretion flow via the ADAF); dashed-dotted red line:  Compton
  cooling from a weak inner disk (25\% of the total accretion flow via
  the disk);
  Upper and lower dotted blue line $T_i$ and $T_e$.}
\end{figure}

\section{Computational results}
\subsection{The effect of irradiation}

We use the computer code described in Liu et al. (2002) including modifications  to the energy flow (Meyer-Hofmeister \& Meyer 2003) and
different components for the Compton process (Meyer-Hofmeister et al. 2005), and an additional component from the inner cool disk. We calculated the coronal structure for a
6$M_ \odot$ black hole, a mean photon energy $h\bar{\nu}$=100 keV for the
radiation from the inner ADAF, and a standard $\alpha$ value of 0.3 for the viscosity.
The evaporation rate varies with the distance from the center, and a
series of calculations for different distances is needed to find the
maximum value of evaporation.

We investigate the influence of heating and cooling of the outer disk
corona by Compton scattering of the additional soft radiation from an
inner disk and the influence of a corresponding reduction of hard
irradiation from the ADAF (with a then reduced mass flow rate).
For the distance where the evaporation rate attains its maximum, Compton scattering produces heating ($q_{\rm{Comp}}<0$), close to the
equatorial plane, and cooling at larger height $z$. This is caused by
the rising electron temperature. This is illustrated in
Fig.\ref{Compton}, which shows $q_{\rm{Comp}}$ for the ``standard
case''for which all matter flows via an ADAF to the center.
The change from heating to cooling
occurs where $4k T_e(z)$ equals $h\bar{\nu}$, in our example at low
vertical height, $z/R= 0.17$. In addition we show how
much an inner disk with a mass flow rate of about 25\% of the total
would contribute. Simultaneously, the ADAF mass flow rate and its
contribution would then be reduced to 75\%, also shown in the figure.
The consequence of a reduced ADAF irradiation for the evaporation
efficiency was already discussed by Liu et al. (2005).

The effect of irradiation on evaporation also depends
on the other contributions to the energy balance such as gains by
friction and vertical advective flow and losses by vertical
heat flow, side wise advective outflow and radiation (for a display see
Meyer et al. 2000a). The hard radiation from the inner ADAF irradiates
the lower coronal layers above the outer disk. Generally additional
heating at low heights tends to support evaporation and raises the
maximal evaporation rate. To prevent a truncation of the outer disk by
complete evaporation, a higher mass flow rate is then required.
This shifts the transition from the low/hard state to the high/soft
state to a higher mass accretion rate and thus a higher luminosity.
Vice versa, a reduction of hard irradiation leads to a lower transition
luminosity.

Fig.\ref{results} shows the maximum of the coronal evaporation rate as
a function of the strength of irradiation from the ADAF. The maximum
determines the transition. In the
standard case of 100\% mass flow in the ADAF the evaporation maximum
is 0.029 $\dot M_{\rm{Edd}}$. As the hard irradiation is decreased, the
maximal evaporation rate also decreases. The evaporation rate also
depends on the mean photon energy of the radiation. For a lower
$h\bar{\nu}$ the maximum evaporation rate is lower, but the dependence on
the strength of the hard irradiation remains similar. Simultaneous
X-ray and $\gamma$-ray observations of Cygnus X-1 in
the hard state by {\it{Ginga}} and OSSE indicate that in this case
the mean photon energy is about $\ge$ 100 keV (Gierli\'nski et al. 1997).

By how much is the hard irradiation decreased? The reduction follows from
two different facts. Firstly it is caused by accretion via the inner
disk instead of the ADAF, secondly by occultation, discussed in the
following paragraph. Since at present the
re-condensation rate into an inner disk has only
been evaluated for low mass flow rates in the ADAF, we do not know
whether the mass flow via an inner disk can be as high as 50\% of the total
accretion rate, suggesting an intermediate state. If a weak inner
disk survived during the quiescent state and the mass
accretion rate increases during the rise to the next outburst, the
mass flow in the inner disk
increases and the disk becomes larger in extent. At the same time, the
inner edge of the outer truncated disk moves inward. The shrinking gap
between the outer and inner disk might finally be filled via re-condensation
or/and inward diffusion of matter from the outer disk when the
critical accretion rate is reached.

\subsection{Occultation and reflection}

The direct hard irradiation is further reduced since an inner
disk can occult a part of the ADAF-filled region. Occultation depends
on the inclination under which the disk appears to a particular
coronal region. The regions relevant for coronal evaporation lie at
low heights, the inclination under which these layers see an inner
disk is generally high, tending to reduce occultation. On the other
hand a larger inner disk leads to more occultation, in the  
limit to 50\%. For example, for an observer at height $z$ above the
mid plane and at a distance of $R=1000R_{\rm S}$, the lower hemisphere
of a spherical optically thin central source of radiation of radius 
$R=6R_ S$, is already fully covered (50\% obscuration) by an
inner disk of radius $R_d=30R_S$ for $z/R > 0.198$, and by an inner
disk of radius
$R_d=100R_S$ for $z/R >0.054$ (this neglects a small ``look
through'' part of order $\frac{3}{8}(\frac{z}{R})^2$ seen through an
inner disk hole of radius 3$R_S$).  Modeling observations of 
GX 339-4 by Miller et al. (2006b) Taam et
al. (2008) obtained an inner disk extension of about 100 $R_S$. The
reduction of hard irradiation by occultation adds to the reduction caused
by letting part of the mass flow via the inner disk instead of via the ADAF.

\begin{figure}[h]
\includegraphics[width=8.8cm]{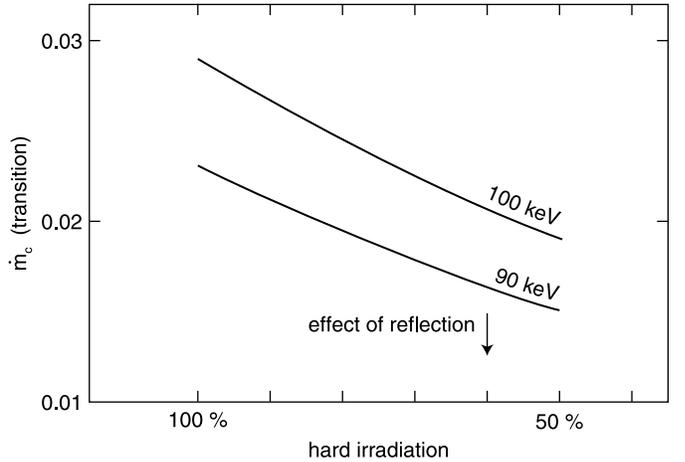}
\caption{\label{results} Critical accretion rate $\dot m_c$ (=$\dot M
  / \dot M_{\rm{Edd}}$) at the hard/soft transition, a measure for the
  transition luminosity, as a function of the strength of irradiation
  ( 100\% irradiation in the case of no inner disk; reduced irradiation for part of the mass
  flow via an inner disk and occultation by the disk). Solid lines:
  $\dot m_c$ for mean photon energy 100 and 90 keV. Reflection leads
  to lower hardness and a further small decrease of transition luminosity. }
   \end{figure}

Part of the hard radiation from the ADAF is Compton scattered at the
disk surface. Thus the back-scattered radiation is reduced in energy
flux and mean photon energy (Lightman \& White 1988). For a 100 keV
mean photon energy one might have a change to 80 keV for the reflected
radiation. If the inner disk is more extended, its temperature drops below about
$10^6$K and absorption of the lower energy photons by the metals
becomes important. This will reduce the soft part in the reflected
radiation and tends to increase its mean photon energy, but at the
same time reduces its total flux. Since only a small amount of
reflected radiation from the
inner disk reaches the outer lower corona due to the large inclination, 
the change of the mean photon energy of irradiation flux caused by reflection 
 is relatively small. Thus, the reflection leads only to 
a small decrease of transition luminosity, though the evaporation rate
depends sensitively on the mean hardness.
Fig.\ref{results} shows how the critical luminosity at transition
(mass flow rate) changes with a reduction of the hard irradiation.
As discussed, the two ways of reduction might add up to 50\%.
Together with the effect of reflection, the hard/soft transition
luminosity may decrease by about a factor of two.

Note that the hard state luminosity seen by an observer is affected by
the presence of an inner disk, as the contribution of the reflected
component is strongly inclination dependent (Esin et al. 1997), 
larger for a lower inclination. 
The inclination of GX 339-4 is not known.
Cowley et al. (2002) suggested that the small emission-line velocity
amplitudes point to a low orbital inclination. From model fits with
Compton reflection, Zdziarski et al. (1998) derived an inclination of
$45^\circ$, Tomsick et al. (2008) argued for an inclination
around $20^\circ$. These values depend on the fits. The dependence on
inclination adds further difficulties to the comparison of the transition
luminosity of different sources.

\section{Comparison with observations}
To compare the luminosities at the hard/soft spectral transition
observed for different sources, one needs to determine
$L/L_{\rm{Edd}}$. To derive this quantity, the distance of the
source, the mass of the black hole, the absorbing column density
towards the source, as well as the instrument response for an
individual observation have to be known, in addition to the (model
dependent) bolometric flux integrated over all energies (for a
discussion see Done \& Gierli\'nski 2003). Therefore, observations of
different outbursts of the same source, as available for GX
339-4, are very valuable.

For GX 339-4 the hard/soft transition during the 1998 and the 2002 outburst
seemed to occur at different luminosities, a factor of three higher
in the latter one than in the former
(Zdziarski et al. 2004). But the observations during the rise in 1998
were scarce. Using their model fits to evaluate the bolometric
luminosity (for a black hole mass 10$M_{\odot}$ and a distance of 8
kpc) the authors found transition luminosities of 0.07$L/L_{\rm{Edd}}$ 
for the hard/soft and  0.025  $L/L_{\rm{Edd}}$ for the soft/hard transitions.
Done \& Gierlinski (2003) show in their Fig.2 the hard/soft transition
at about 0.04 $L/L_{\rm{Edd}}$ by adopting a black hole mass of
6$M_{\odot}$ and a distance of 4 kpc. The difference is mainly due to the
different black hole mass and distance assumed.

 In the detailed investigation of Dunn et
al. (2008), four outbursts are considered, in 1998, 2002, 2004
and 2007. Comparing the two best sampled outbursts in 2002 and 2004 it
is clear that the hard/soft transition in 2004 occurs at an intensity
 three to four times lower than that in 2002. During the 2007 outburst the
transition is similar to that in 2002. On the other hand, the transitions during
decline back to the hard state were very similar in the outbursts of
2002/2003 and 2004. From the summed spectra (Dunn et al. 2008,
Fig.5) no conclusion on a hardness change due to an inner disk in 2004
is possible. Further investigations of GX 339-4 included an analysis
of the {\it{RXTE}} and {\it{INTEGRAL}} high-energy
observations of a hard/soft transition in 2004 by Belloni et
al. (2006), as well as {\it{RXTE}} and {\it{INTEGRAL}} observations of the
transition in 2007 by Del Santo et al. (2009) and {\it{INTEGRAL}} and
{\it{XMM-Newton}} observations for the same transition by
 Caballiero-Garc\'ia et al. (2009).

Now {\it{XMM-Newton}} spectra clearly revealed a cool
accretion disk component for GX 339-4 during the rise to the
outburst in 2004 (Miller et al. 2006b, Reis et al. 2008). This was the
rise to outburst where the transition occurred at a luminosity three
times lower than during the outburst in 2002, supporting the
suggestion that an inner disk affects the state transition
luminosity. The effects discussed above can add up
to a mass accretion rate at transition lower by a factor of 2, to
 be compared with
the observed difference of a factor of three in luminosity. The exact
value depends on the particular parameters describing the system. We
note that the luminosity might depend non-linearly on the mass accretion rate.

For neutron star transients the situation is less clear. If one expects the
same accretion flow geometry as in black hole binaries, an additional inner
disk could also provide enhanced soft radiation and Compton cooling,
together with reduced hard irradiation. During their shorter outburst
cycles, the quiescence is usually better covered by observations,
providing information on how deep the luminosity decreases and whether
an inner disk might survive. Reig et al. (2004) analyzed five
outbursts of Aquila X-1 and investigated
spectral states. In their Table 1, the maximal intensities of hard radiation
during the rise are given for different outbursts, showing a
difference of transition luminosity of about 10\% for outburst 4 and 5.

Hard/soft transitions of Aquila X-1 were also described by Yu \&
Dolence (2007). In the context of our considerations it seems interesting
that during a
very weak outbursts in July 2001 (after maybe not too low an intensity during
quiescence), a hard/soft transition occurred at a low luminosity and,
a short time
after this, the transition back to the hard state at the same
luminosity, without hysteresis.
In our theoretical picture an inner disk can survive if the mass
accretion rate does not decrease too much during quiescence. This
might also allow us to understand why for Cyg X-1, with luminosity
changes of only a factor 3-4, no hysteresis is observed (Zdziarski et
al. 2002). The inner disk probably never disappears completely.
It is of interest that Miller et al. (2002b) found an inner disk during the
intermediate state of Cygnus X-1 in January 2001.

\section {The hardness intensity diagram}

\begin{figure}[h]
\includegraphics[width=8.cm]{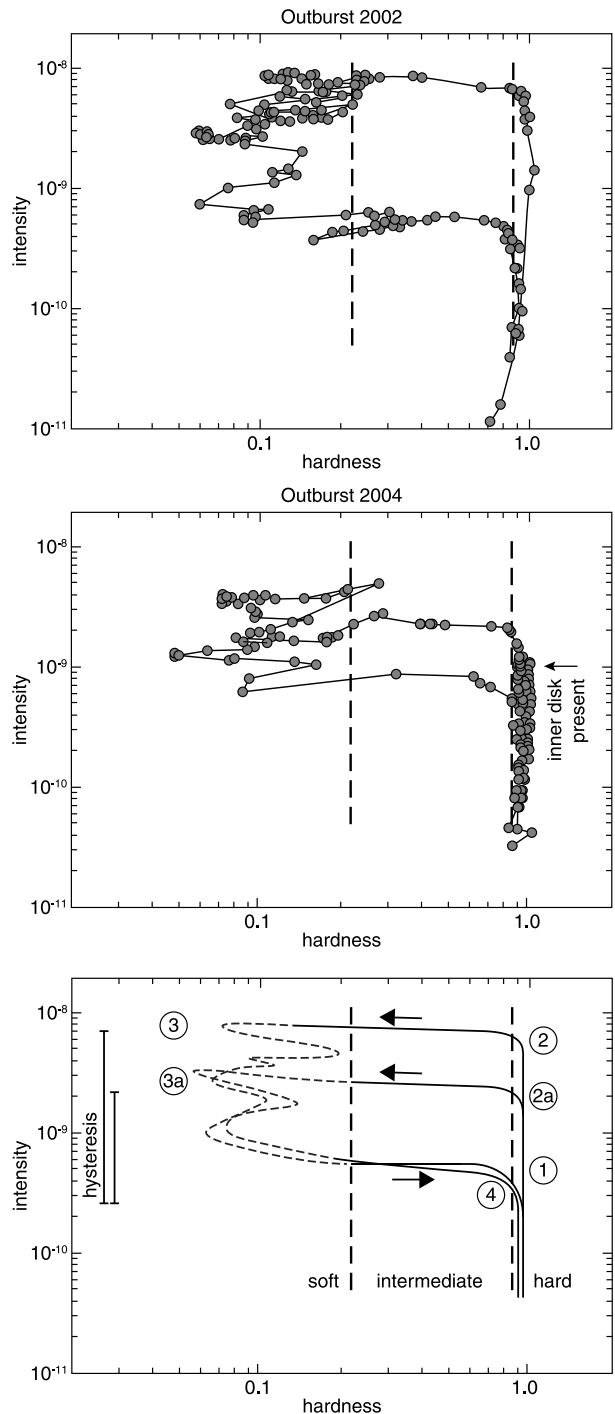}
\caption{\label{HID}
         HID of GX 339-4: Upper and middle diagram from Dunn et
         al. (2008, Fig.4), outbursts in 2002 and 2004,
         intensity: 3-10keV flux, hardness: flux ratio 6-10keV/3-6keV,
         long-dashed vertical lines mark state transitions.
         Bottom: track expected from theory, (1) rise to outburst,
         (2),(2a) transition to the soft state, (3),(3a) peak
         luminosity, (4) decline from outburst, transition to hard state.}
   \end{figure}

In Fig.\ref{HID}, upper and middle diagram, we show hardness intensity
diagrams for the
2002 and 2004 outburst of GX 339-4 from Dunn et al. (2008, Fig.4). The
hard/soft transition luminosity is clearly different, lower during
the rise to outburst 2004 where the inner disk was observed (Miller et
al. 2006b).

In the bottom diagram we show the diagram derived from our theoretical
picture of accretion geometry. The position in the HID changes with the mass
accretion rate during the outburst cycle. For the rise to
outburst (low accretion rate, state(1) in Fig.\ref{HID}) the outer
disk is truncated and the
inner region filled by an ADAF. We expect a hard spectrum determined
by the ADAF near the center. During the rise to higher luminosity and
increasing intensity, the hardness is almost constant. This branch in the HID
always has a similar appearance, as occurs for other sources. When the
accretion rate becomes higher than the maximal evaporation rate, the ADAF in the
inner region is replaced by a disk (state(2)).
This is usually a rapid change, with
only little increase in intensity, resulting in a horizontal movement
to small hardness values in the HID (the hardness changes in the HID can be
estimated from the color-color diagrams in the investigation of
Done \& Gierli\'nski 2003). At what intensity the hard/soft
transition occurs depends, as we show  here, on
the presence of an inner disk.

As soon as the outer disk reaches inward to the last stable orbit,
the spectrum is soft; during the further run of the outburst, mainly
the intensity changes until a peak value (state (3) or (3a)) is reached. Small
variations of hardness, variations of the 6-10keV flux, might result from
more or less mass flow in a weak corona. Depending on the increase and
decrease of the
luminosity, the path in the HID might appear somewhat irregular,
different from outburst to outburst. We therefore describe this
episode with dashed lines in the theoretical HID.

After the soft state, the return to the hard state occurs when the mass
accretion rate decreases below a critical value (state(4)). Note that during the
soft state the coronal structure and the evaporation are affected by soft
irradiation from the disk reaching inward to the last stable
orbit. This is different from the hard irradiation from the inner ADAF
during the rise to outburst. The different irradiation causes the well-known
hysteresis in the transition luminosity (Meyer-Hofmeister et
al. 2005).

The track in the HID is basically the same for all outbursts although a few
special features during the outburst can change its appearance. (1) In
outbursts where the luminosity increases to values much higher than at
the hard/soft state transition, these data appear high up in the left upper
area of the HID.
(2) In some outbursts the soft state is not reached, and sometimes only an
intermediate spectral state is reached, as discussed by Brocksopp et
al. (2004). Well known examples are  V404 Cyg (Oosterbroek et
al. 1997), XTE J1118+480 (Remillard et al. 2000, Hynes et al. 2000,
Revnivtsev et al. 2000) and XTE1550-564 (Sturner \& Shrader 2005). Very
recently a so-called ``failed outburst'' was observed for 
H 17432-322 (Capitanio et al. 2009), there the parts of the HID related 
to soft spectra are not
reached during the outburst. Such behavior can be understood as
related to accumulation of only small amounts of mass, which preferably happens in
binaries with short orbital periods, that is, a small accretion disk
(Meyer-Hofmeister 2004).

\section{Discussion}

Other effects that might play a role in the hard/soft transition
luminosity are the following.

The vertical structure of the corona depends on the viscosity. We used
$\alpha$=0.3 for the viscosity parameter, supported by modeling of X-ray binary
spectra (Esin et al. 1997). A higher viscosity parameter leads to higher
evaporation efficiency and a higher transition luminosity (Liu et al. 2002,
Qiao \& Liu 2009). If the same higher viscosity parameter also holds for
the transition back to the hard state that transition would also occur
at higher luminosity, which was not observed, at least for the well
studied outbursts of GX 339-4 in 2002 and 2004. Viscosity changes
could be related to changes of the magnetic flux present at the
beginning of the outburst. But such fluxes would have to change until
the decline.

A change of the radiation efficiency due to an increasing mass flow
via an inner disk would lead to a higher luminosity, corresponding to a
certain accretion rate at transition. The resulting luminosity would
be slightly higher than for constant $\eta=0.05$. Based on earlier
investigations, Narayan \& McClintock (2008) had pointed out that there is
little difference between the two efficiencies for an ADAF or disk
accretion for mass flow rates close to spectral transition.

\section{Conclusions}
Recent observations for GX 339-4 and J1753.5-0127 by Miller et
al. (2006a, 2006b), Tomsick et al. (2008) and Reis et al. (2008)
confirm the presence of an inner disk during the rise and decline of
outbursts. Further sources of observations also indicate a possible inner
disk. Theoretical investigation showed that these inner disks can
survive if re-condensation of matter from the ADAF allows us to sustain
the disk embedded in the ADAF. This is possible as long as the
luminosity does not drop below $ 10^{-3}-10^{-4} L_{\rm{Edd}}$ in
quiescence (Liu et al. 2006, Meyer et al. 2007).

The observations for several sources seem to show that the hard/soft
transition does not always occur at the same luminosity.
A comparison of transition luminosities of different sources is made
difficult as the values obtained depend on several parameters
taken for the individual systems. Thus it is of special interest
to compare the transitions for the same source, best sampled for
GX 339-4. The hard/soft transition
during the 2004 outburst occurred at a luminosity three times lower than
that in the 2002 outburst. We investigate whether the presence of an
inner disk can account for these differences. Such an inner disk was
documented for the rise to the outburst in 2004.

 Our investigation shows how the hard/soft spectral
transition would occur at lower luminosity if an inner disk
had remained from the previous outburst. The transition luminosity is
affected by the reduction of hard irradiation from the ADAF seen by
the corona, due to a decreased accretion via the ADAF and occultation
of the innermost region by the inner disk. A secondary effect is
caused by a decrease of the mean photon energy as direct and reflected
radiation is mixed.
It would be interesting to study a
source shortly before the state transition in the rise to outburst.
For very low inclination systems we would predict a
decrease of the hardness in the spectrum.

\begin{acknowledgements}
We thank Marat Gilfanov for fruitful discussions.
We are grateful for information which the anonymous referee provided
 on additional cases of soft components in low/hard state systems
which led to a significant improvement of Table 1. BFL acknowledges
supports from the National Natural Science Foundation of China (grants
10533050 and 10773028) and from the National Basic Research Program of
China-973 Program 2009CB824800.

\end{acknowledgements}

{}


\begin{thebibliography}{}
\bibitem{} Abramowicz, A., Chen, X., Kato, S. et al., 1995, ApJ 438,
  L37
\bibitem{} Belloni, T., Parolin, I., Del Santo, M. et al. 2006, MNRAS 367, 1113
\bibitem{} Bradley, C.K., Hynes, R.I., Kong, A.K.H. 2007, ApJ 667, 427
\bibitem{} Brocksopp, C., Bandyopadhyay R. M., \& Fender, R.P. 2004
  New Astronomy 9, 249
\bibitem{} Caballero-Garc\`ia, M.D., Miller, J.M., D\`iaz Trigo, M. et
  al. 2009, ApJ 692, 1339
\bibitem{} Capitanio, f., Belloni, T., Del Santo et al. 2009, MNRAS
  398, 1194
\bibitem{} Corbel, S.,Tomsick, J.A.,\& Kaaret, P. 2006, ApJ 636, 971
\bibitem{} Cowley, A.P., Schmidtke, P.C., Hutchings, J.B. et al. 2002,
  ApJ 123, 1741
\bibitem{} Del Santo, M., Bazzano, A., Zdziarski, A.A. et al. 2005,
  A\&A 433, 617
\bibitem{} Del Santo, M.Belloni, T.M., Homan, J. et al. 2009, 392, 992
\bibitem{} Done, C. \& Gierlinsli, M. 2003, MNRAS, 342, 1041
\bibitem{} Dunn, R.J.H., Fender, R.P., K\"ording, E.G. et al. 2008,
  MNRAS 387, 545
\bibitem{} Esin, A.A. 1997, ApJ 482, 400
\bibitem{} Esin, A.A., McClintock, J.E., \& Narayan, R. 1997, ApJ 489,
  865
\bibitem{} Gierli\'nski, M., Zdziarski, A.A., Done, C. et al. 1997,
  MNRAS 288, 958
\bibitem{} Hynes, R.I., Mauche, C.W., Haswell, C.A. et al. 2000, ApJ
  539, L37
\bibitem{} Ho, L. C. 2008, ARA\&A 46, 475
\bibitem{} Kalemci, E., Tomsick, J.A., Buxton, M.M. et al. 2005,
  ApJ 622, 508
\bibitem{} in't Zand, J.J.M., Markwardt, C.B., Bazzano, A. et al. 2002a, A\&A 390, 597
\bibitem{} in't Zand, J.J.M., Miller, J.M., Oosterbroek,T. et al. 2002b, A\&A 394, 553
\bibitem{} La Palombara, N. \& Mereghetti, S. 2005, A\&A 430, L53 
\bibitem{} Lightman, A.P., \& White, T.R. 1988, ApJ 335, 57
\bibitem{} Liu, B. F., Mineshige, S., Meyer F. et al. 2002, ApJ 575,
  117
\bibitem{} Liu, B. F., Meyer, F., \& Meyer-Hofmeister, E. 2005, A\&A
  454, L9
\bibitem{} Liu, B. F., Meyer, F., \& Meyer-Hofmeister, E. 2006, A\&A
  442, 555
\bibitem{} Liu, B. F., Taam, R.E., Meyer-Hofmeister, E. et al. 2007,
  ApJ 671, 695
\bibitem{} McClintock, J.E., Haswell, C.A.,Garcia, M.R. et al. 2001,
  ApJ 555, 477
\bibitem{} McClintock, J.E., \& Remillard, R.A., 2006, In: Compact
Stellar X-ray Sources, eds. W.H.G. Lewin, M. van der Klis,
Cambridge Astrophysics Series, No. 39, Cambridge University Press,
2006, p. 157
\bibitem{} Meyer, F., Liu, B.F., \& Meyer-Hofmeister, E. 2000a, A\&A,
361,175
\bibitem{} Meyer, F., Liu, B.F., \& Meyer-Hofmeister, E. 2000b, A\&A,
354, L67
\bibitem{} Meyer, F., Liu, B.F., \& Meyer-Hofmeister, E. 2007, A\&A
  463, 1
\bibitem{} Meyer-Hofmeister, E. \& Meyer, F. 2003, A\&A 402, 1013
\bibitem{} Meyer-Hofmeister, E. 2004, A\&A 423, 321
\bibitem{} Meyer-Hofmeister, E., Liu, B.F., Meyer, F. 2005, A\& A 432,
  181
\bibitem{} Miller, J.M., 2007, ARA\&A, 45, 441
\bibitem{} Miller, J.M., Fabian, A.C., Wijnands, R. et al. 2002a, ApJ
  570, L69
\bibitem{} Miller, J.M., Fabian, A.C., Wijnands, R. et al. 2002b, ApJ
  578, 348
\bibitem{} Miller, J.M., Fabian, A.C., in 't Zand, J.J.M. et
  al. 2002c, ApJ 577, L15
\bibitem{} Miller, J.M., Homan, J., \& Miniutti, G. 2006a, ApJ 652,
  L113
\bibitem{} Miller, J.M., Homan, J., Steeghs, D. et al. 2006b, ApJ 653,
  525
\bibitem{} Miller, J.M., Reynolds, C.S., Fabian, A.C. et al. 2009, ApJ
  697, 900
\bibitem{} Miyamoto, S., Kitamoto, S. Hayashida, K. et al. 1995, ApJ
  442, L13
\bibitem{} Narayan, R., \& Yi, I. 1994, ApJ 428, L13
\bibitem{} Narayan, R., \& Yi, I. 1995a,  ApJ 444, 231
\bibitem{} Narayan, R., \& Yi, I. 1995b,  ApJ 452, 710
\bibitem{} Narayan, R., \& Yi, \& Mahadevan, R. 1995c, Nature 374, 623 
\bibitem{} Narayan, R., McClintock, J.E. \&  Yi, I. 1996, ApJ 457,
  821
\bibitem{} Narayan, R., Barret, D., \& McClintock, J.E. 1997, ApJ
  482, 448
\bibitem{} Narayan, R., Mahadevan, R., \& Quataert, E. 1998, In: The
Theory of Black Hole Accretion Discs, eds. M.A. Abramowicz et al.,
Cambridge Univ. Press, p. 48
\bibitem{} Narayan, R. \& McClintock, J.E. 2008, New Astronomy Reviews
  51, 733
\bibitem{} Nowak, A.M. 2006, arXiv:astro-ph/0611909
\bibitem{} Oosterbroek, T., van der Klis, M., van Paradijs, J. et
  al. 1997, A\&A 321, 776 
\bibitem{} Park, S.Q., Miller, J.M., McClintock, J.E. et al. 2004, ApJ
  610, 378
\bibitem{} Pottschmidt, K., Chernyakova, M., Zdziarski, A.A. et
  al. 2006, A\&A 452, 285
\bibitem{} Qiao, E., \& Liu, B.F. 2009, PASJ 61, 403
\bibitem{} Reig, P., van Straaten, S. \& van der Klis, M. 2004 ApJ
  602, 918
\bibitem{} Reis, R.C., Fabian, A.C., Ross, R. et al. 2008, MNRAS 392,
  992
\bibitem{} Reis, R.C., Miller, J.M., \& Fabian, A.C. 2009, MNRAS 395, L52
\bibitem{} Remillard, R.A., Morgan, E.Smith, D. et al. 2000, IAU
  Circ. 7389
\bibitem{} Remillard, R.A., \& McClintock, J.E. 2006, ARA \& A 44, 49
\bibitem{} Revnivtsev, M.G., Sunyaev, R., \& Borozdin, K. 2000, A\&A
  361, L37
\bibitem{} Rossi, S., Homan, J., Miller, J.M. et al. 2005, MNRAS 360, 763
\bibitem{} Rykoff, E.S., Miller, J.M., Steeghs, D. et al. 2007, ApJ
  666,1129
\bibitem{} Sturner, S.J., \& Shrader, C.R. 2005, ApJ 625, 923
\bibitem{} Taam, R.E., Liu, B.F., Meyer, F. et al. 2008 ApJ 688, 527
\bibitem{} Tomsick, J.A., Kalemci, E., \& Kaaret, P. 2004, ApJ 601, 439
\bibitem{} Tomsick, J.A., Kalemci, E., Kaaret, P. et al. 2008, ApJ
  680, 593
\bibitem{} Yu, W. \& Dolence, J. 2007, ApJ 667, 1043
\bibitem{} Yuan, F. 2007, ASP Conference Series, Vol. 373, p. 95
\bibitem{} Yuan, F. \& Narayan, R. 2004, ApJ 612, 724
\bibitem{} Zdziarski, A.A., Poutanen, J., Miko{\l}ajewska, J. et
  al. 1998, MNRAS 301, 435
\bibitem{} Zdziarski, A.A., Poutanen, J. Paciesas, W.S. et al. 2002,
  ApJ 578, 357
\bibitem{} Zdziarski, A.A., Gierli\'nski, M., Miko{\l}ajewska, J. et
  al. 2004, MNRAS 351, 791

\end{thebibliography}
\end{document}